\documentclass{ctr}

\usepackage{ctrfont}
\usepackage{natbib}

\usepackage{url}
\usepackage{amsmath}
\usepackage{amssymb}

\usepackage[dvips]{graphicx}
\usepackage{psfrag}
\usepackage{epsfig}

\newcommand{\Figref}[1]{Figure~\ref{#1}}
\newcommand{\Figrefs}[1]{Figures~\ref{#1}}

\newcommand{\Tabref}[1]{Table~\ref{#1}}

\newcommand{\Eqnref}[1]{Eq.~\eqref{#1}}
\newcommand{\Eqnrefs}[1]{Eqs.~\eqref{#1}}

\usepackage{color}
\usepackage{enumitem}

\title{Inter-scale energy transfer in turbulence from the viewpoint of subfilter scales}
\shorttitle{Inter-scale energy transfer in turbulence}
\author{J.~I. Cardesa\footnote{School of Aeronautics, Technical University of Madrid, Spain}
  \footnote{Present address: Institut de M\'ecanique des Fluides de Toulouse, INPT-CNRS-UPS, France}
  \and A. Lozano-Dur\'an}
\shortauthor{Cardesa \& Lozano-Dur\'an}

\begin{document}


\maketitle


\section{Motivation and objectives}

Understanding the cascading process of kinetic energy in turbulent
flows from large-scale motions to smaller scales is critical in
modeling strategies for geophysical and industrial flows. The
phenomenological explanation of the transfer of energy across scales
was conceptualized, first by \cite{Richardson1922} and later by
\cite{Obukhov1941}, as interactions among energy eddies of
different sizes. The picture was completed in the classical paper by
\cite{Kolmogorov1941}, and since then, many detailed investigations
have greatly advanced our understanding of the energy cascade in high-
Reynolds-number ($Re$) turbulent flows. Notwithstanding these efforts,
the cascading process remains one of the most challenging problems in
turbulence due to its non-linear and multiscale nature. In the present
work, we investigate the inter-scale energy transfer (ISET) in turbulence
from the viewpoint of subfilter scales (SFS). We show that this new
perspective provides an amenable framework to characterize the energy
transfer.

Attempts to unravel the mechanisms behind the cascade have relied on
varying but complementary physical rationales.  A classic explanation
is given in terms of vortex stretching acting across scales
\citep{Leung2012, Goto2017, LozanoHolznerJFM2016, Motoori2019}. Other
approaches have directly tested Richardson's idea in
terms of eddy breakdown \citep{lozano_jimenez_JFM2014}, or energy
transfer among eddies at different scales \citep{Cardesa_Science}.
The advent of multifractal representations of the cascade
\citep{Meneveau1991, Frisch1991a, Frisch1996, Mandelbrot1999,
  Jimenez2000, Yang2017} and wavelet methods \citep{Farge1992,
  Schneider2010} have also opened new avenues for analyzing the
statistical properties of the energy transfer across scales.
 
A distinction should be made between the different approaches discussed
above in terms of (i) tools, which may include wavelets
\citep{Meneveau1991}, coherent structures
\citep{lozano_jimenez_JFM2014}, space-time correlations
\citep{Wan_Meneveau_Eyink_PoF010}, filtering of the equations
\citep{Piomelli1981}, and so forth, and (ii) quantities, which may
include source/sink terms in the energy \citep{Aoyama} or the
enstrophy \citep{Leung2012, LozanoHolznerJFM2016} equation, vortex
stretching and strain self-amplification \citep{Betchov1956,
Tsinober2009}, the kinetic energy itself \citep{Cardesa_Science},
and so on. What may appear as a purely semantic debate is in fact
an attempt to tackle the problem of how to look at the energy
cascade. Here, we are concerned with choosing the most appropriate
quantity to characterize the energy cascade, since tools only differ
in how they handle this quantity.
 
The starting point for many studies of the energy cascade is the
decomposition of the flow into resolved and unresolved (or subfilter)
scales, inherent to high-$Re$ turbulence. This partition of
the flow has been extensively used both for physical understanding
of the energy cascade and for its reduced-order modeling. In
the latter, accurate predictions of the flow require a faithful
interpretation of the subfilter physics and of its interaction with
the resolved field. In this context, a set of ubiquitous terms are
given by the SFS stresses,
\begin{equation}
\tau_{ij} = \overline{u_{i}u_{j}} - \overline{u}_{i}\overline{u}_{j}, \label{eqn:sgs_stress_def}
\end{equation}
appearing in the filtered momentum equations, written here in
incompressible form
\begin{equation}
\partial_{t}\overline{u}_{i} + \overline{u}_{j}\partial_{j}\overline{u}_{i} = -\partial_{i}\overline{p} - 
\partial_{j} \tau_{ij} +\nu\partial_{j}\partial_{j}\overline{u}_{i}+\overline{F}_{i}. \label{eqn:filt_momentum}
\end{equation}
We follow the convention of decomposing the total velocity $u_{i}$
into $u_{i}= \overline{u}_{i} + u'_{i}$, where
\begin{equation} \label{eqn:decomp_def}
    \overline{u}_{i}\left( \boldsymbol{x} \right) =
  \int u_{i}\left( \boldsymbol{x}-\boldsymbol{r}\right) G\left( \boldsymbol{r} \right)\mathrm{d}\boldsymbol{r}
\end{equation}
is the resolved velocity defined with respect to some filter
$G(\boldsymbol{r})$. The spatial Cartesian coordinates are
$\boldsymbol{x} \equiv (x_1,x_2,x_3)$ (and occasionally
$\boldsymbol{x}\equiv (x,y,z)$), $F_{i}$ is a forcing term, and
$\overline{F}_{i}$ is its filtered counterpart. The kinematic
viscosity of the flow is $\nu$, and the pressure scaled by a constant density
is $p$.
The term
$P=\overline{S}_{ij}\tau_{ij}$ is the ISET,
and it appears with opposite sign in the evolution equations of
two energy definitions: $\overline{u}_{i}\overline{u}_{i}$ and $\tau_{ii}$.

To model $\partial_{j}\tau_{ij}$
in \Eqnref{eqn:filt_momentum}, it is essential that $P$ be consistent with
phenomenological aspects of the energy cascade. As a result, modelers are often
concerned with the properties of $P$ when feeding physics into
their large-eddy simulation (LES) models. We argue, however, that studying the physics of the
energy cascade through term $P$ can be significantly improved upon. We
describe the issues associated with the use of $P$ as a marker for
ISET, and propose an alternative ISET
term more suitable to study the spatial structure of energy transfer
across scales.

This brief is organized as follows. In Section \ref{sec:equations} we
introduce the filter-scale energy equations, discuss the properties of
the sources and sinks of kinetic energy, and identify our target features
concerning an optimal ISET term. A similar analysis is performed in
Section \ref{sec:equations_res} for the kinetic energy associated with
the subfilter velocities, and a new energy transfer term $T$ is
introduced. Comparisons between $T$ and $P$ for different flows and
for various filters are offered in Section \ref{sec:comparisons_flow} and
Section \ref{sec:comparisons_filter}. Finally, we present our
conclusions in Section \ref{sec:conclusions}.

\section{Equations for the filter-scale kinetic energy}
\label{sec:equations}

\subsection{Subfilter-scale energy transfer and alternative formulations}
\label{subsec:alternative}

The traditional ISET term $P$ is found in the equation for the
kinetic energy of the filtered velocities,
\begin{eqnarray}
      \left( \partial_{t}+ \overline{u}_{j}\partial_{j} \right)  \frac{1}{2}\overline{u}_{i}\overline{u}_{i} &= -\partial_{j} \left(\overline{u}_{j}\overline{p} + \overline{u}_{i}\tau_{ij} -2\nu \overline{u}_{i} \overline{S}_{ij} \right) - 2\nu \overline{S}_{ij}\overline{S}_{ij}
      +\underbrace{\overline{S}_{ij}\tau_{ij}} +\overline{u}_{i}\overline{F}_{i}.   \label{eqn:kin_e_filt}\\
\nonumber & \hspace{5.7cm} P
\end{eqnarray}
$P$ is commonly interpreted as the rate of transfer of
kinetic energy from the filtered motions to the residual motions
\citep{Pope}, i.e., energy transport in scale. 
The frequent justification for this meaning stems from
the fact that $P$ is outside the divergence, so that it cannot be 
energy transport in space. While there is evidence and agreement that, for a given
flow field, the spatial distribution and statistical properties of $P$
depend heavily on the filter used \citep{Piomelli1981, Aoyama, cardesaPoF2015},
it is less commonly acknowledged that there are
alternative formulations of \Eqnref{eqn:kin_e_filt}. In alternative formulations,
the resulting term outside the divergence could be assigned the same interpretation
as that of $P$, since whatever is outside of the divergence (and is non-viscous)
should be transport in scale. For instance,
\begin{eqnarray}
      \left( \partial_{t}+ \overline{u}_{j}\partial_{j} \right)  \frac{1}{2}\overline{u}_{i}\overline{u}_{i} &= -\partial_{j} \left(\overline{u}_{j}\overline{p} -2\nu \overline{u}_{i} \overline{S}_{ij} \right) - 2\nu \overline{S}_{ij}\overline{S}_{ij}
      +\underbrace{\overline{u}_{i}\partial_{j}\left(\tau_{ij}\right)}+\overline{u}_{i}\overline{F}_{i}.   \label{eqn:kin_e_filt2}\\
\nonumber & \hspace{4.6cm} P_{2}
\end{eqnarray}
$P_{2}$ and $P$ have the same spatially averaged value in the absence
of boundary fluxes, yet their pointwise values differs
significantly. Since LES models benefit from information on the
physics of what occurs at each grid point, arguing that $P_{2}$ and
$P$ only differ by an amount that cancels out on the average does not
settle an important issue: what is the pointwise ISET of the flow?
The question is central to LES models that attempt to handle the
presence of the so-called backscatter, which is a local property of
$P$ \citep{Piomelli1981,Cerutti_PoF1998}. The question also highlights the need
for refined criteria when assigning a meaning to a term in an equation.

Another concerning aspect of $P$ in \Eqnref{eqn:kin_e_filt} is that
the term $\overline{u}_{i}\tau_{ij}$ inside the divergence of
\Eqnref{eqn:kin_e_filt} can be expanded as
\begin{equation}
  \overline{u}_{i}\tau_{ij}=\overline{u}_{i}\left(\overline{\overline{u}_{i}\overline{u}_{j}} + 
\overline{\overline{u}_{i}u'_{j}} +
\overline{\overline{u}_{j}u'_{i}} +
  \overline{u'_{i}u'_{j}} - \overline{u}_{i}\overline{u}_{j}\right).
\end{equation}
The terms in parentheses include both filter-scale and SFS
interactions, so that a product of $\tau_{ij}$ with $\overline{u}_{i}$
involves inter-scale coupling by necessity. Hence, $P$ does not
contain all coupling between filter scales and SFS in the
right-hand side of \Eqnref{eqn:kin_e_filt}. In view of these
observations, it is difficult to assign to $P$ the role of causing all
energy changes stemming from filter-subfilter interactions. This
illustrates the challenge of trying to distinguish between terms
transporting energy entirely in space or purely across scales.

\subsection{Galilean invariance}

An additional consideration is that of Galilean invariance. Briefly,
the Galilean transformation is defined as
\begin{equation}
\hat{x}_{i}=x_{i}-V_{i}t; \hspace{1cm} \hat{t}=t; \hspace{1cm} \hat{u}_{i}=u_{i}-V_{i}, \label{eqn:gali_trans}
\end{equation}
where $V_{i}$ is the constant velocity of the frame of reference. We
provide in the Appendix some identities related to \Eqnref{eqn:gali_trans} 
used on filtered and residual velocity
components. Applying the transformations to the momentum equations, we find that
the rate of change on the left-hand side preserves both its form and
its value under Galilean transformations,
\begin{equation}
\left( \partial_{t} + u_{j}\partial_{j} \right) u_{i} =  \left[ 
\hat{\partial}_{t}-V_{j}\hat{\partial}_{j}+\left(\hat{u}_{j}+V_{j}\right) \hat{\partial}_{j}\right]\left( \hat{u}_{i}+V_{i} \right)=
\left( \hat{\partial}_{t}+\hat{u}_{j}\hat{\partial}_{j} \right) \hat{u}_{i}. \label{eqn:Navier_gali_trans}
\end{equation}
The same applies to the filtered momentum equations.

The situation contrasts starkly with the transformed kinetic
energy equation. Applying the transformations to \Eqnrefs{eqn:kin_e_filt} and
(\ref{eqn:kin_e_filt2}), we find that the left-hand side becomes
\begin{equation}
\left( \partial_{t} + \overline{u}_{j}\partial_{j} \right) \frac{1}{2}\overline{u}_{i}\overline{u}_{i} = V_{i}\left( \hat{\partial}_{t} + \overline{\hat{u}}_{j}\hat{\partial}_{j} \right)\overline{\hat{u}}_{i} + \left( \hat{\partial}_{t} + \overline{\hat{u}}_{j}\hat{\partial}_{j} \right)
\frac{1}{2}\overline{\hat{u}}_{i}\overline{\hat{u}}_{i}. \label{eqn:gali_largescale_en}
\end{equation}
In the first term of the right-hand side of
\Eqnref{eqn:gali_largescale_en}, $V_{i}$ multiplies the left-hand side
of the (filtered) momentum equations.  This term cancels out with an
equivalent term on the right-hand side of the transformed
\Eqnref{eqn:kin_e_filt} or \Eqnref{eqn:kin_e_filt2}, so that overall, 
\Eqnrefs{eqn:kin_e_filt} and (\ref{eqn:kin_e_filt2}) preserve their form
and are thus Galilean invariant.
However, the rates
of change of energy observed on
different Galilean frames of reference will differ in value, even if
the law those changes obey appears to be the same.

To illustrate the difference, we present the following thought
experiment.  Let the rate of change $(\partial_{t} +
\overline{u}_{j}\partial_{j})\overline{u}_{i}\overline{u}_{i}/2$ at a
given point and instant be $q$ on the static frame of reference. Then,
according to \Eqnref{eqn:gali_largescale_en}, it is possible to choose
a $V_{i}$ such that the first term on the right-hand side of
\Eqnref{eqn:gali_largescale_en} equals $q$.  In that Galilean frame of
reference, the rate of change of energy given by the last term in
\Eqnref{eqn:gali_largescale_en} would be zero.

The argument above shows how a Galilean-invariant equation needs not
lead to rates of change which are themselves Galilean invariant. In addition,
individual terms in a Galilean-invariant equation need not be Galilean invariant.
For example, it is straightforward to prove that $P$ is Galilean invariant
while $P_{2}$ is not, yet both appear in Galilean-invariant equations.
Hence, $P$ is a better ISET term candidate than $P_{2}$ in terms of
independence from the frame of reference.

The implications of the discussion above lead us to conclude that
a preliminary step in choosing a suitable form of the ISET term is the
choice of the appropriate equation. Indeed, it is desirable to study
the terms in an equation which preserves identical rates of change on
different Galilean frames of reference, as is the case for the
momentum equation.  We also seek, if possible, an equation where terms
are individually Galilean invariant. A drawback of not doing so
appears when looking at the pointwise value of individual terms. If
the latter are not Galilean invariant term by term, then the relative
importance of, say, a divergence term compared to an ISET term will
depend on the frame of reference, which is unsatisfactory from a
fundamental point of view --- even though the sum of all terms on both
sides of the equation balances out. The considerations regarding
Galilean invariance of individual terms are similar to those
concerning improved expressions for the SFS stress models
\citep{Speziale1, Germano1} discussed in the past.

\section{Equations for the subfilter-scale kinetic energy}
\label{sec:equations_res}

It was argued above that studying the kinetic energy transfer from the
point of view of the filter scales entails a series of drawbacks
which are intrinsic to the formulation of the equation itself. Several
prerequisites are identified in the previous section concerning the
target ISET term. We summarize them below. In essence, we are aiming
to obtain an ISET term that:
\begin{enumerate} [label=\alph*)~,leftmargin=1.5\parindent]
\item belongs to an equation that conserves the value of the rate of
  change of energy under Galilean transformations,
\item belongs to the right-hand side of an equation where each product is
  individually Galilean invariant,
\item avoids filter-subfilter coupling inside the spatial flux term, and
\item avoids filter-only or subfilter-only coupling.
\end{enumerate}

The decomposition between filter scales and SFS has triggered
the consideration of several kinetic energy expressions. One of the
most extended list of expressions is perhaps that presented in
Section 3.3.1 of \cite{SagautLESbook}. Nonetheless, little or no
consideration has been paid to the evolution equation for the kinetic
energy of the subfilter velocities.  It can be shown that the equation
for the kinetic energy of the subfilter (residual) scales preserves
the value of the rate of change under Galilean
transformations. Various equations can be derived depending on the
particular arrangement of terms on the right-hand side,
\begin{eqnarray}
\left( \partial_{t}+ \overline{u}_{j}\partial_{j}\right) \frac{1}{2}u'_{i}u'_{i} &=&  -\partial_{j} \left(
        u'_{j}p'-2\nu u'_{i}S'_{ij}
        \right) - 2\nu S'_{ij}S'_{ij} \nonumber \\
        &+& \underbrace{u'_{i}\partial_{j}\left(\tau_{ij}\right)-u'_{i}u'_{j}\overline{S}_{ij}-u'_{i}u'_{j}S'_{ij}}_{T_{a}}+u'_{i}F'_{i} \label{eqn:kin_e_resA} \\
\left( \partial_{t}+ \overline{u}_{j}\partial_{j}\right) \frac{1}{2}u'_{i}u'_{i} &=&  -\partial_{j} \left(
        u'_{j}p'-2\nu u'_{i}S'_{ij}-u'_{i}\tau_{ij}
        \right) - 2\nu S'_{ij}S'_{ij} \nonumber     \\
        &+& \underbrace{-S'_{ij}\tau_{ij}-u'_{i}u'_{j}\overline{S}_{ij}-u'_{i}u'_{j}S'_{ij}}_{T_b}+u'_{i}F'_{i}\label{eqn:kin_e_resB}\\
\left( \partial_{t}+ \overline{u}_{j}\partial_{j}\right) \frac{1}{2}u'_{i}u'_{i} &=&  -\partial_{j} \left(
        u'_{j}p'-2\nu u'_{i}S'_{ij}+\frac{1}{2} u'_{i}u'_{i}u'_{j}
        \right) - 2\nu S'_{ij}S'_{ij} \nonumber \\
        &+& \underbrace{u'_{i}\partial_{j}\left(\tau_{ij}\right)-u'_{i}u'_{j}\overline{S}_{ij}}_{T_c}+u'_{i}F'_{i}\label{eqn:kin_e_resC}\\
\left( \partial_{t}+ \overline{u}_{j}\partial_{j}\right) \frac{1}{2}u'_{i}u'_{i} &=&  -\partial_{j} \left(
        u'_{j}p'-2\nu u'_{i}S'_{ij}-u'_{i}\tau_{ij}+\frac{1}{2} u'_{i}u'_{i}u'_{j}
        \right) - 2\nu S'_{ij}S'_{ij} \nonumber \\
        &+& \underbrace{S'_{ij}\tau_{ij}-u'_{i}u'_{j}\overline{S}_{ij}}_{T_d}+u'_{i}F'_{i}.\label{eqn:kin_e_resD}\\
\nonumber 
\end{eqnarray}

As stated previously, the motivation for looking at these equations is
that the left-hand side complies with prerequisite a), which
\Eqnref{eqn:kin_e_filt} fails to do. The term
$\overline{u}_{i}\partial_{j}(u'_{i}u'_{j})$ appears in alternative
ways of writing the right-hand side of
\Eqnrefs{eqn:kin_e_resA}--(\ref{eqn:kin_e_resD}), which is
inconsistent with b), so we do not consider those expressions. Visual
inspection of \Eqnrefs{eqn:kin_e_resB} and (\ref{eqn:kin_e_resD})
reveals that some filter-subfilter coupling remains inside the
divergence through the $u'_{ij}\tau_{ij}$ term, which goes against
c). Choosing between \Eqnrefs{eqn:kin_e_resA} and
(\ref{eqn:kin_e_resC}) relies on compliance with d), which
\Eqnref{eqn:kin_e_resC} achieves to a larger extent by avoiding the
presence of $u'_{i}u'_{j}S'_{ij}$ inside the ISET term.  Expanding
$\tau_{ij}$ within $T_{c}$, we find the product
$u'_{i}\partial_{j}\overline{u'_{i}u'_{j}}$. While its interpretation
as a subfilter-only term is arguable, it appears in the equation
which comes closest to complying with all four established
prerequisites, and therefore we choose $T_{c}$ as our ISET term and
relabel it as
\begin{equation}
T=u'_{i}\partial_{j}\left(\tau_{ij}\right)-u'_{i}u'_{j}\overline{S}_{ij}. \label{eqn:def_T}
\end{equation}
When comparing \Eqnrefs{eqn:kin_e_resC} and (\ref{eqn:kin_e_filt}), one should not
forget that
\begin{equation}
u_{i}u_{i}=\overline{u}_{i}\overline{u}_{i}+2\overline{u}_{i}u'_{i}+u'_{i}u'_{i}. \label{eqn:sum_energies}
\end{equation}
There are thus a wealth of additional energy equations related to the evolution of
$\overline{u}_{i}u'_{i}$,
which invariably fail to comply with a). It is tempting to assume that,
by construction, we have removed any dependence on the large scales of the flow by
looking at the evolution
of $u'_{i}u'_{i}$ instead of $\overline{u}_{i}u'_{i}$ or $\overline{u}_{i}\overline{u}_{i}$.
It is not so, however, since the evolution equations considered above relate to rates of
change as perceived by an observer traveling with velocity $\overline{u}_{j}$, which
preserves the large-scale information. In addition, the energy gained or lost through $T$
invetably comes from or goes to the large scales, so in principle there is no reason to
believe we are enforcing a removal of large-scale dependence.

Some statistical properties of terms $T_a$, $T_b$, $T_c$, and $T_d$
were studied by \cite{CardesaBertinoro2014} for one flow, confirming that $T_c$
featured the most desirable properties of all four candidates in terms of acting
as the small-scale counterpart of $P$.  
In the next section, we compare the views of ISET obtained
through $T$ and $P$ for several flows.
Before doing so, we note in passing that compliance with a) would also 
hold for the evolution equation of $\tau_{ii}$. However, we discard that
equation for failing to comply with b) and c), as detailed in the Appendix.
%
\section{Comparison between $T$ and $P$ for different flows}
\label{sec:comparisons_flow}

We compute $T$ and $P$ in three different flows obtained by direct
numerical simulation (DNS), namely, homogeneous shear turbulence
(HST), homogeneous isotropic turbulence (HIT), and a plane
turbulent channel flow (CH4K). The details of the three simulations
can be found in \Tabref{tab:sim_details}.  The large scales of the
three flows selected differ significantly from one another, so that we
can benchmark the degree of dependence of $T$ and $P$ on the large
scales.

All three flows have been filtered with an isotropic low-pass Gaussian
filter defined as
\begin{equation}
\tilde{G}(\boldsymbol{k})=\exp\left[-\left(r\boldsymbol{k} \right)^{2}/40 \right] \label{eqn:filt_k_def}
\end{equation}
in Fourier space, where $\boldsymbol{k}$ is the wave number and
$r$ is the filter width. The filtering was applied to the Fourier
modes of the velocity components of HIT, as well as in
the streamwise and spanwise directions of HST and CH4K.  In HST,
filtering along $y$ was implemented by convolving the velocity with
the real-space transform of $G\left(y\right) =
f\exp[-6y^{2}/r^{2}]$, where $f$ is a constant chosen to meet the
normalization condition. In CH4K, the filtering in the wall-normal
direction was carried out following the procedure described by
\cite{LozanoHolznerJFM2016}.  We used only values of $T$ and $P$
extracted from the plane at the channel centerline, where the flow
inhomogeneity is the weakest.  The width of the largest filter used is
$672\eta$, where $\eta$ is the Kolmogorov length scale based on the
dissipation at the channel centerline. The wall-to-wall distance is
$1062\eta$, and the widest filter extends from the centerline down to
$y/h=0.37$ or $y^{+}=1533$ from both walls. This is close to the edge of the
log-law region \citep{LozanoPoF2014}, so that most filter widths
characterize essentially the core of the channel.
\begin{table}
  \begin{center}
    \begin{tabular}{c c c c c}
      ~Case~ & \hspace{0.5cm} $Re_{\lambda}$ & \hspace{0.5cm} $N_{x} \times N_{y} \times N_{z}$ & \hspace{0.5cm} $(L_{x} \times L_{y} \times L_{z})/\eta$ & \hspace{0.5cm} $L_{o}/\eta$ \\
      HST & \hspace{0.5cm} 107 & \hspace{0.5cm} $768\times512\times255$ & \hspace{0.5cm} $1117 \times 745 \times 372$ & \hspace{0.5cm} 267 \\
      CH4K & \hspace{0.5cm} 161 & \hspace{0.5cm} $2048 \times 541 \times 2047$& \hspace{0.5cm}$3337\times 1062 \times 1668$ & \hspace{0.5cm} 1397\\
      HIT  & \hspace{0.5cm} 236 & \hspace{0.5cm} $512^{3}$ & \hspace{0.5cm} $1011^{3}$ & \hspace{0.5cm} 876 \\

    \end{tabular}
  \end{center}
  \caption{Details of the three simulations used: homogeneous shear
    turbulence (HST), a plane channel flow at friction-based Reynolds
    number $Re_{\tau}=4179$ (CH4K), and homogeneous isotropic
    turbulence (HIT). All flows are incompressible and
    statistically steady.  $N_{i}$ and $L_{i}$ are the number of
    degrees of freedom and the domain size in directions $i=x,y,$ and $z$,
    which for the shear flows HST and CH4K correspond to streamwise,
    vertical, and spanwise directions, respectively.  The Kolmogorov
    length scale $\eta$ is given by $(\nu^{3}/\epsilon)^{1/4}$, where
    $\nu$ is the kinematic viscosity and $\epsilon$ the mean rate of
    kinetic energy dissipation. For CH4K, $\eta$ is that measured at
    the channel centerline. For HST and CH4K, the Reynolds number
    based on the Taylor microscale is defined as
    $Re_{\lambda}=K^{2}[5/(3\nu\epsilon)]^{1/2}$, where $K^{2}$ is
    twice the kinetic energy of the velocity field fluctuating with
    respect to the time-averaged, $y$-dependent streamwise mean
    velocity. For HIT, $Re_{\lambda}$ is computed in the standard way,
    while $K^{2}$ is twice the total kinetic energy.
    $L_{o}=K^{3}/\epsilon$ is a length scale characterizing the large
    scales. For more details on the simulations, CH4K is case M4200
    in \cite{LozanoPoF2014}, while HST and HIT are documented by
    \cite{cardesaPoF2015}.
  \label{tab:sim_details}}
\end{table}
%

\subsection{Net amount of forward and backward energy cascade}

\begin{figure}
  \vspace{0.2cm}
  \begin{center}
    \includegraphics[width=0.4\textwidth]{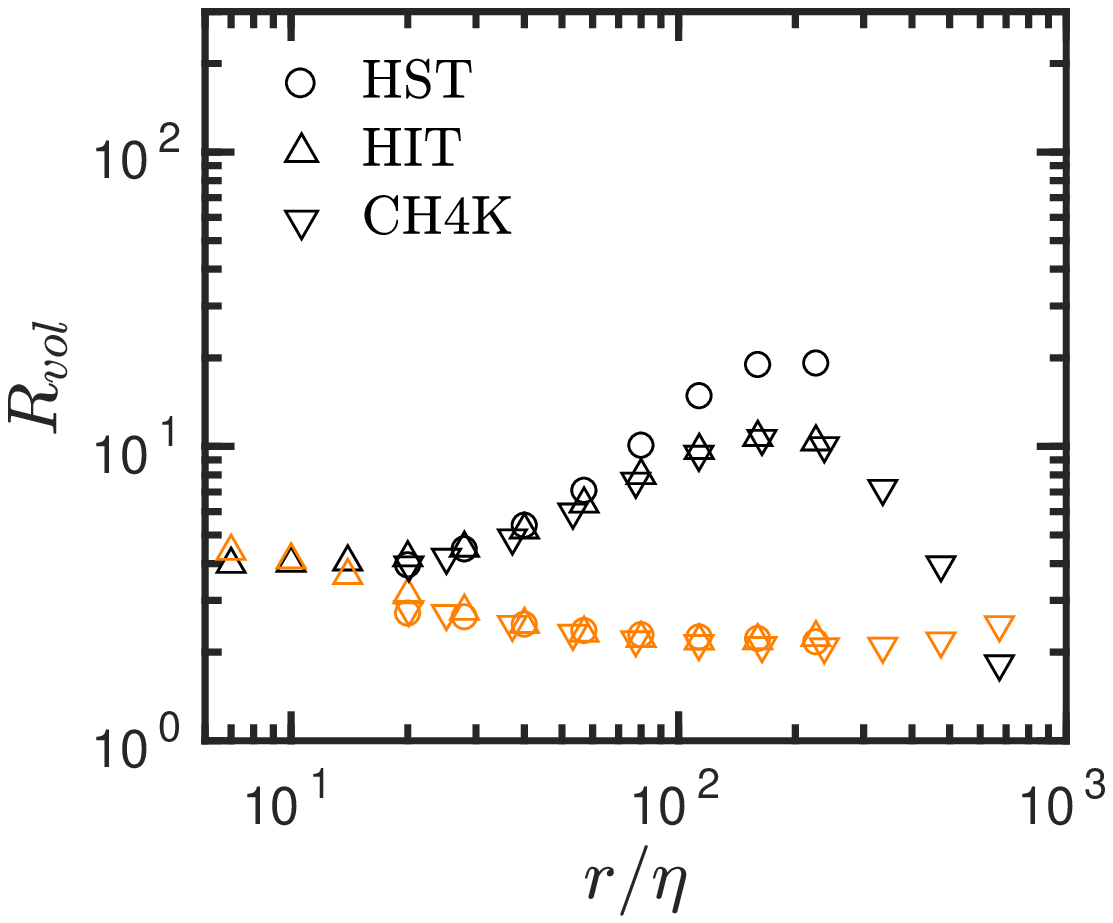}\hspace{.5cm}
    \includegraphics[width=0.4\textwidth]{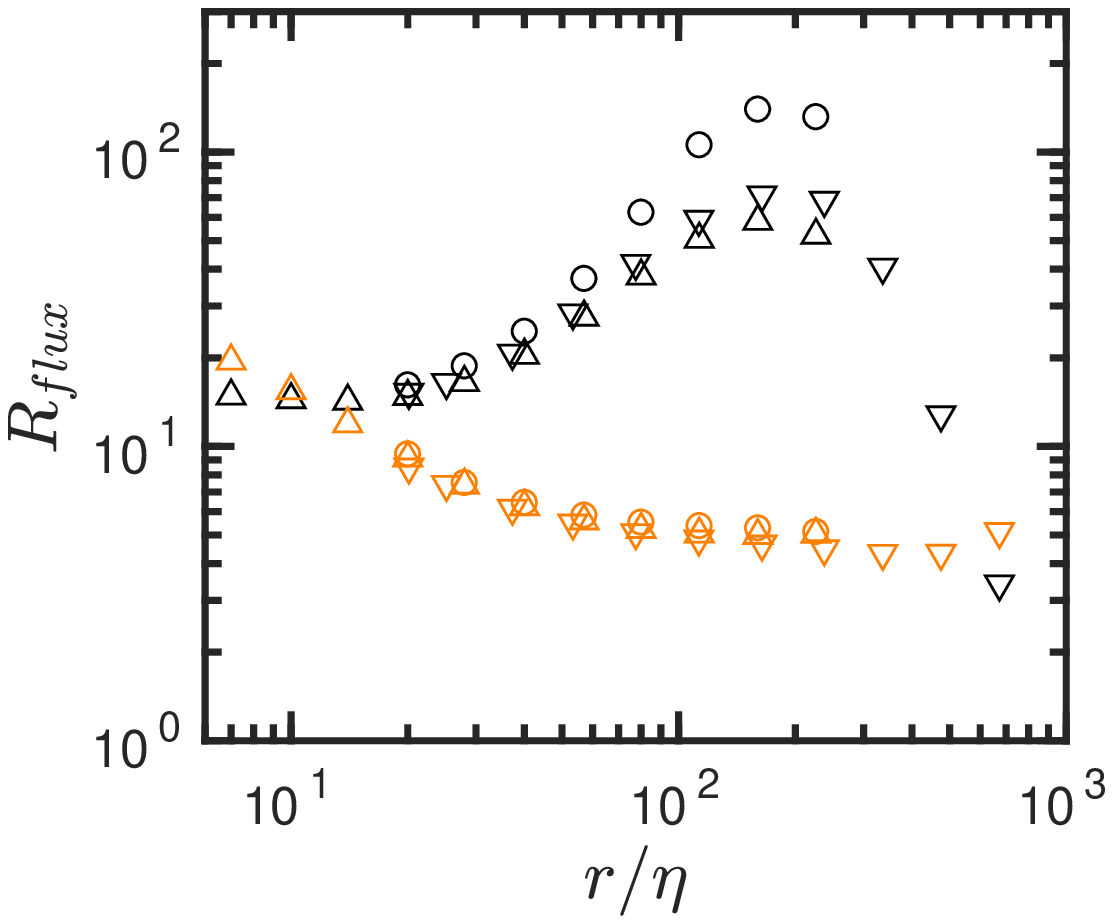}\\
    \hspace{-5.1cm} (a) \hspace{5.5cm} (b)
    \caption{Ratios of direct (forward) over inverse (backward)
      cascade based on $T$ (orange) and $P$ (black). For $T$, forward and
      backward cascade correspond to $T>0$ and $T<0$,
      respectively. $P<0$ and $P>0$ are taken as forward and backward
      cascade, respectively.  (a) (flow volume forward) / (flow volume
      backward), (b) (energy flux forward) / (energy flux
      backward).\label{fig:cascade_ratios}}
  \end{center}
\end{figure}

The ratio between the volumes of flow experiencing forward to
backward ISET is shown in \Figref{fig:cascade_ratios}(a). The spatial
predominance of the forward cascade is perceived through both ISET
terms $T$ and $P$. Ratios based on $T$ are lower than for $P$, except for the
narrow filter widths in the viscous range, where they converge to similar values.
The most striking difference between the picture of the cascade obtained with both ISET 
terms is that ratios based on $P$ exhibit a stronger $r$ dependence than those
based on $T$. It is best illustrated by the drop at large $r$ in the ratio of the
CH4K case when based on $P$ and not on $T$. In addition, the ratio in the HST case differs
significantly from the
other two flows for larger $r$ when computed with $P$ and not so with $T$. 
The largest filter width in the HST
case is $227\eta$, which is perhaps close enough to the spanwise extent of the box
($372\eta$) to have an impact on the statistics. But whatever the reason for the
divergence from the other two flows, the ratio for HST based on $T$ is not impacted,
while that based on $P$ is. It appears that 
$T$ filters out differences across flows and across $r$, providing a more homogeneous
view of the
cascade. Very similar conclusions are reached by looking at the ratios
of forward to backward ISET in terms of actual energy flux in
\Figref{fig:cascade_ratios}(b). The main difference between
\Figref{fig:cascade_ratios}(a) and \Figref{fig:cascade_ratios}(b) is
that the ratios of forward to backward cascade are higher
when quantified with the energy fluxes rather than with flow
volumes, and this is clear for both $T$ and $P$.
%

\subsection{Structure of forward/backward-cascading transfer eddies}

\begin{figure}
  \vspace{0.1cm}
  \begin{center}
    \includegraphics[width=0.48\textwidth]{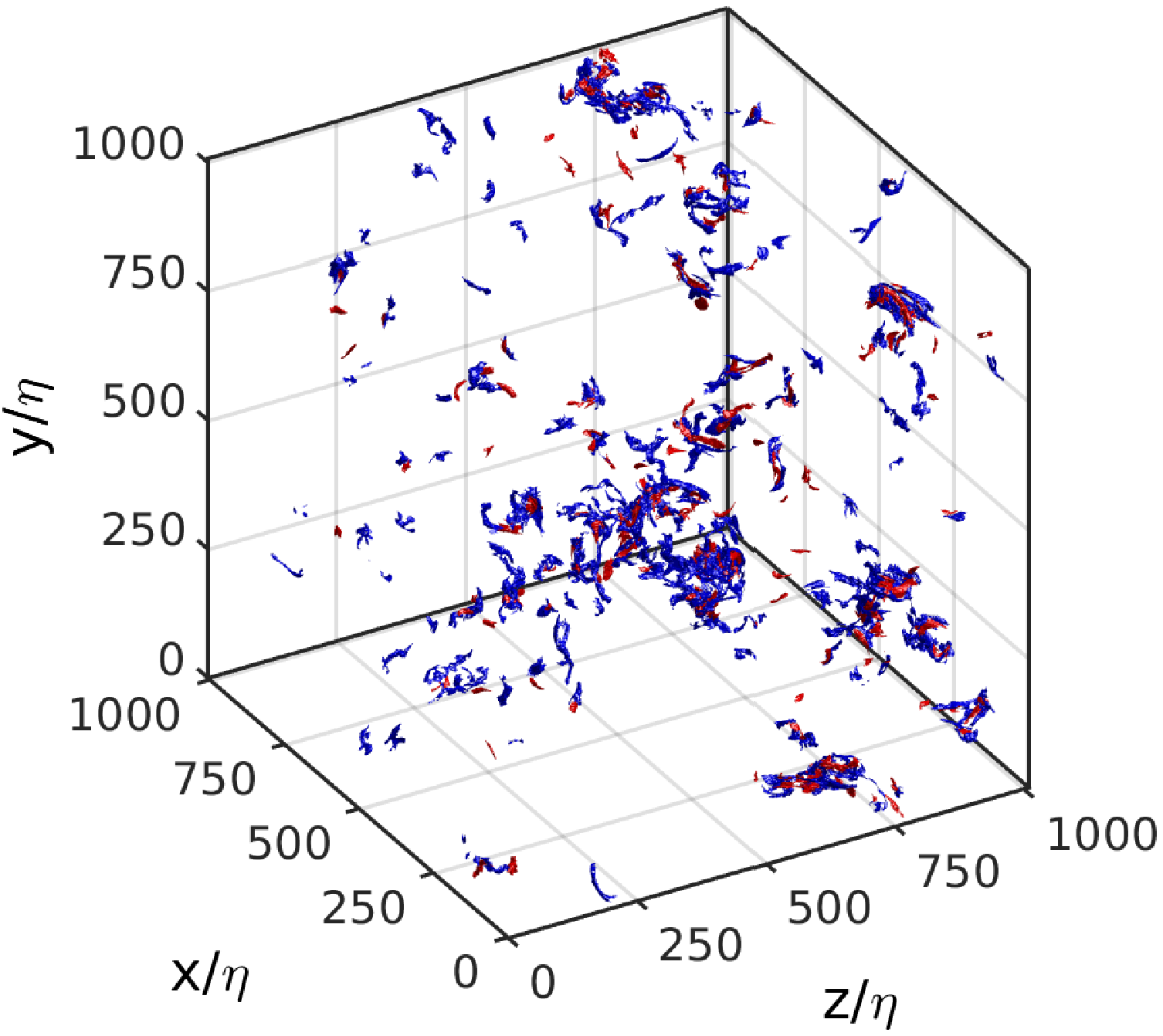}\hspace{0cm}
    \includegraphics[width=0.48\textwidth]{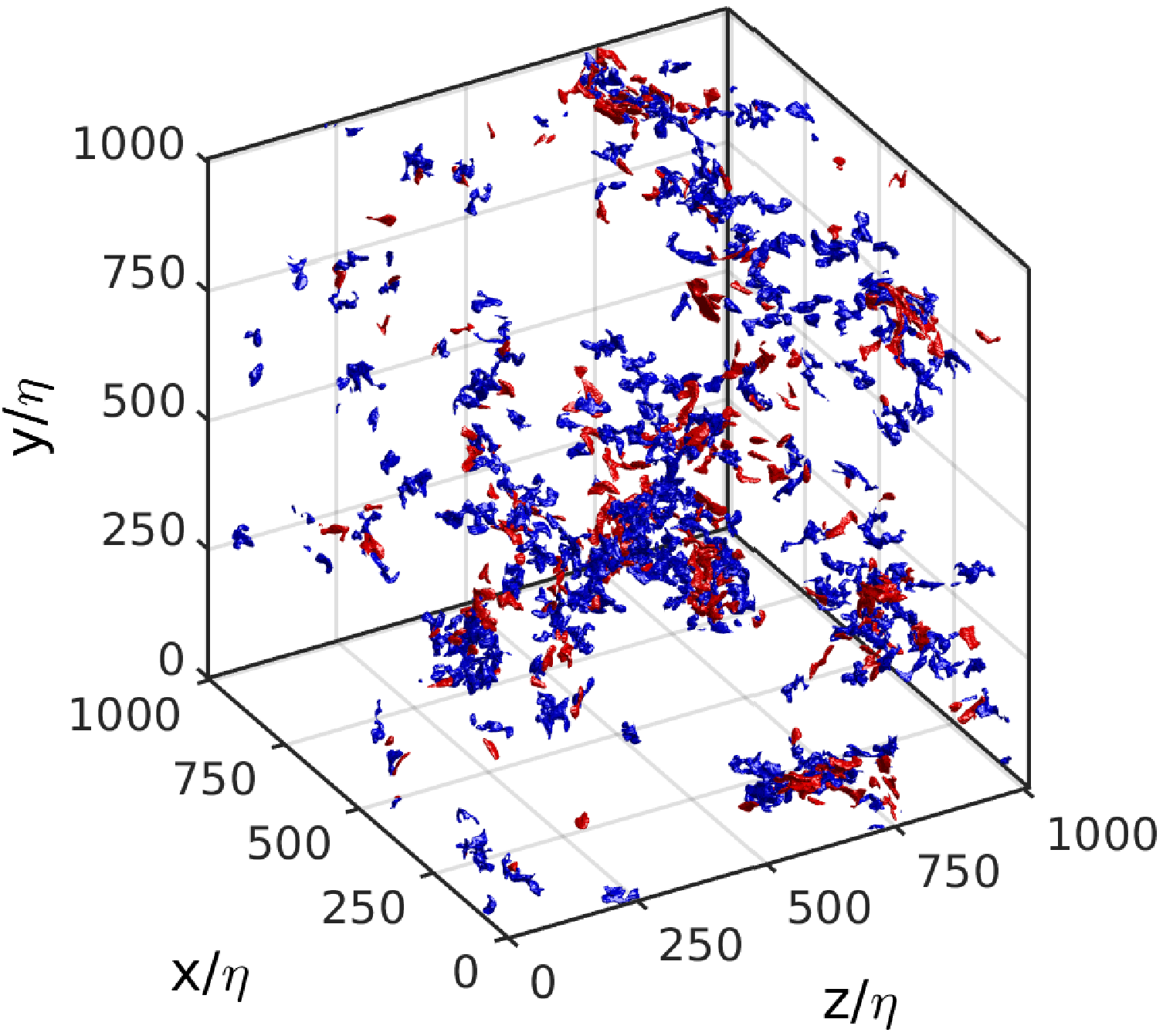}\\
    \hspace{-5.1cm} (a) \hspace{5.5cm} (b)\\
    \vspace{-.07cm}
    \includegraphics[width=0.48\textwidth]{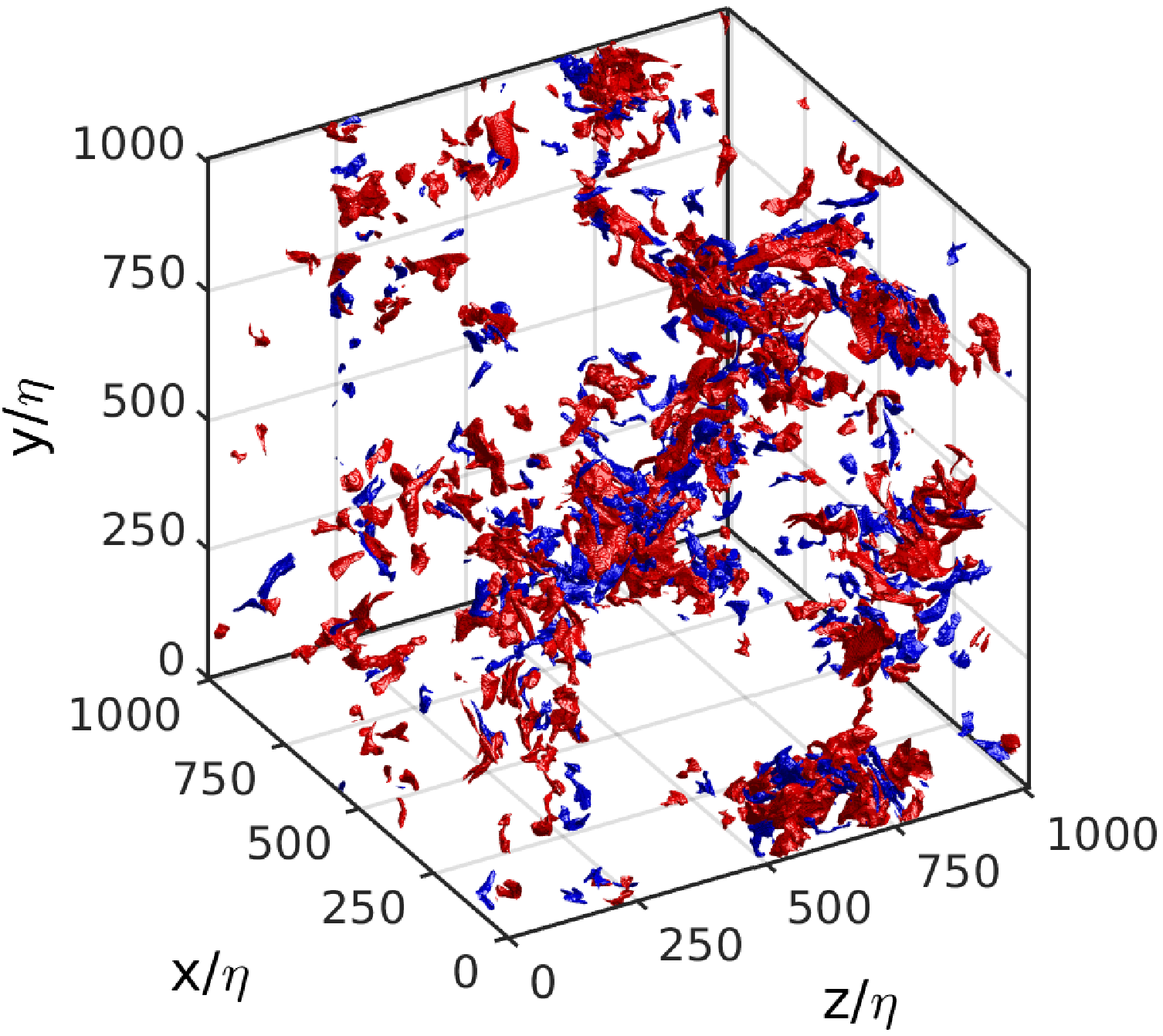}\hspace{0cm}
    \includegraphics[width=0.48\textwidth]{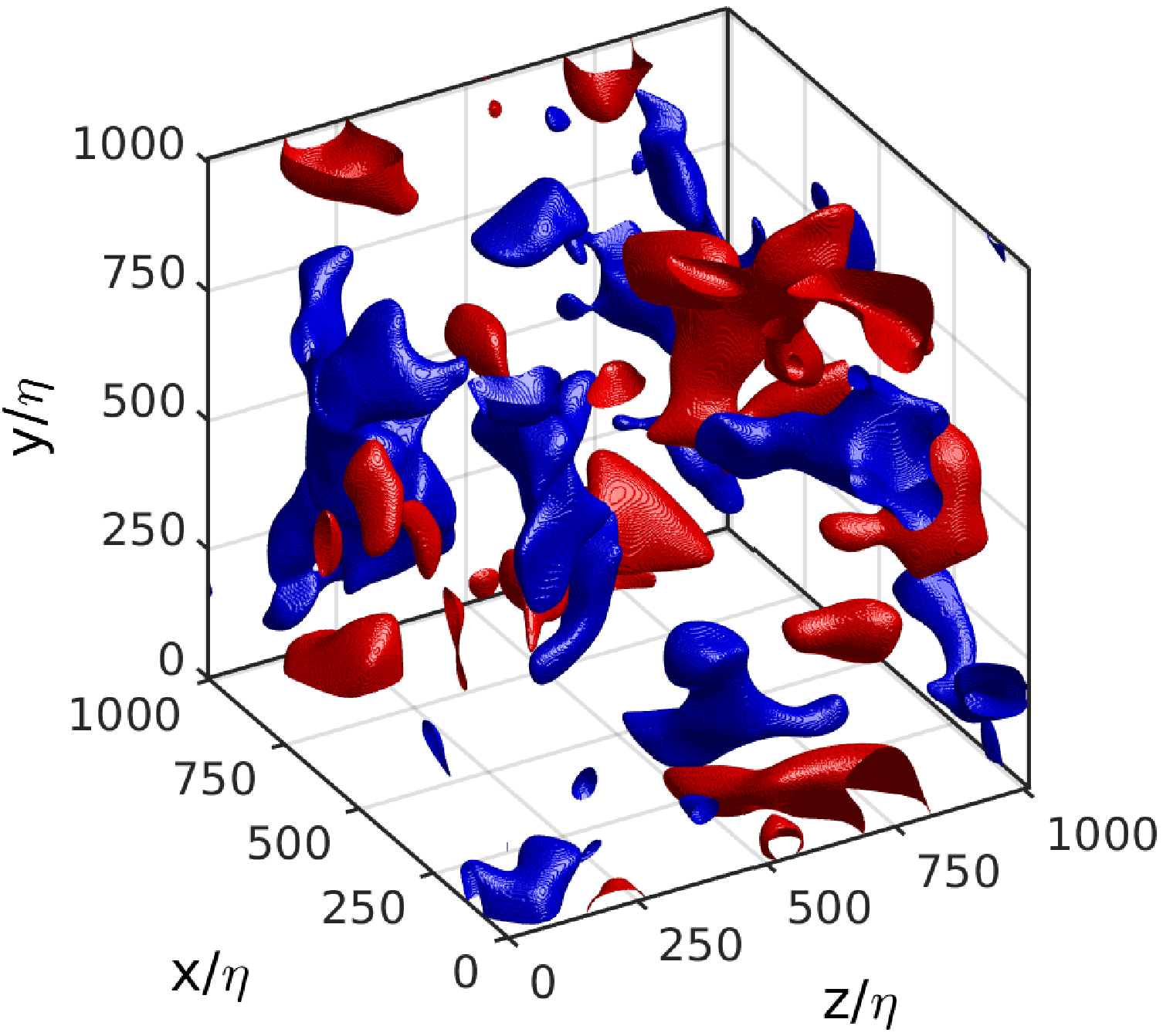}\\
    \hspace{-5.1cm} (c) \hspace{5.5cm} (d)
    \caption{Case HIT: objects of intense $|T|$ (left) and intense
      $|P|$ (right). (a) and (b): $r=14\eta$.  (c) and (d):
      $r=160\eta$. Red objects contain forward cascade ($P<0$ or
      $T>0$), blue objects contain backward cascade ($P>0$ or
      $T<0$). \label{fig:structures}}
  \end{center}
\end{figure}

We now focus on case HIT and study the properties of the scalar
field $T>\alpha^+$, where $\alpha^+>0$ is a thresholding
parameter. For each $r$, we select a value of $\alpha^+$ such that the
rate of change of the total volume enclosed by $T>\alpha^+$ decreases
most with increasing threshold \citep[see the percolation theory
  by][]{Kesten1982}. We repeat the process for the field $T<\alpha^-$
with $\alpha^-<0$. The resulting structures of intense $|T|$
(collectively denoting $T>\alpha^+$ and $T<\alpha^-$) are plotted in
\Figref{fig:structures}(a) for $r=14\eta$ and in
\Figref{fig:structures}(c) for $r=160\eta$. The same steps are
followed for $P$, shown in \Figrefs{fig:structures}(b,d). 
The visualizations reveal that $|T|$ and
$|P|$ structures obtained for $r=14\eta$ bear clear similarities,
while those obtained for $r=160\eta$ do not. This is in agreement with
the fact that, in \Figref{fig:cascade_ratios}, the disparity between
the cascade ratios computed for $T$ and $P$ is considerably smaller for
filter widths within the viscous range than for filter widths above
$100\eta$. Furthermore, by comparing the structures of intense $|T|$
at $r=14\eta$ and $r=160\eta$, one is able to observe similarities
between \Figrefs{fig:structures}(a,c),
consistent with the similar events that have simply grown in scale. It
is challenging, however, to see any similarities between
\Figrefs{fig:structures}(b) and \ref{fig:structures}(d), so that the
picture that emerges through $P$ is extremely dependent upon the
filter width $r$ --- at least, much more so than upon $T$. Again, this is
consistent with \Figref{fig:cascade_ratios}, in which
cascade ratios based on $T$ are significantly less $r$ dependent
than those based on $P$.

\begin{figure}
  \begin{center}
    \includegraphics[height=4.2cm]{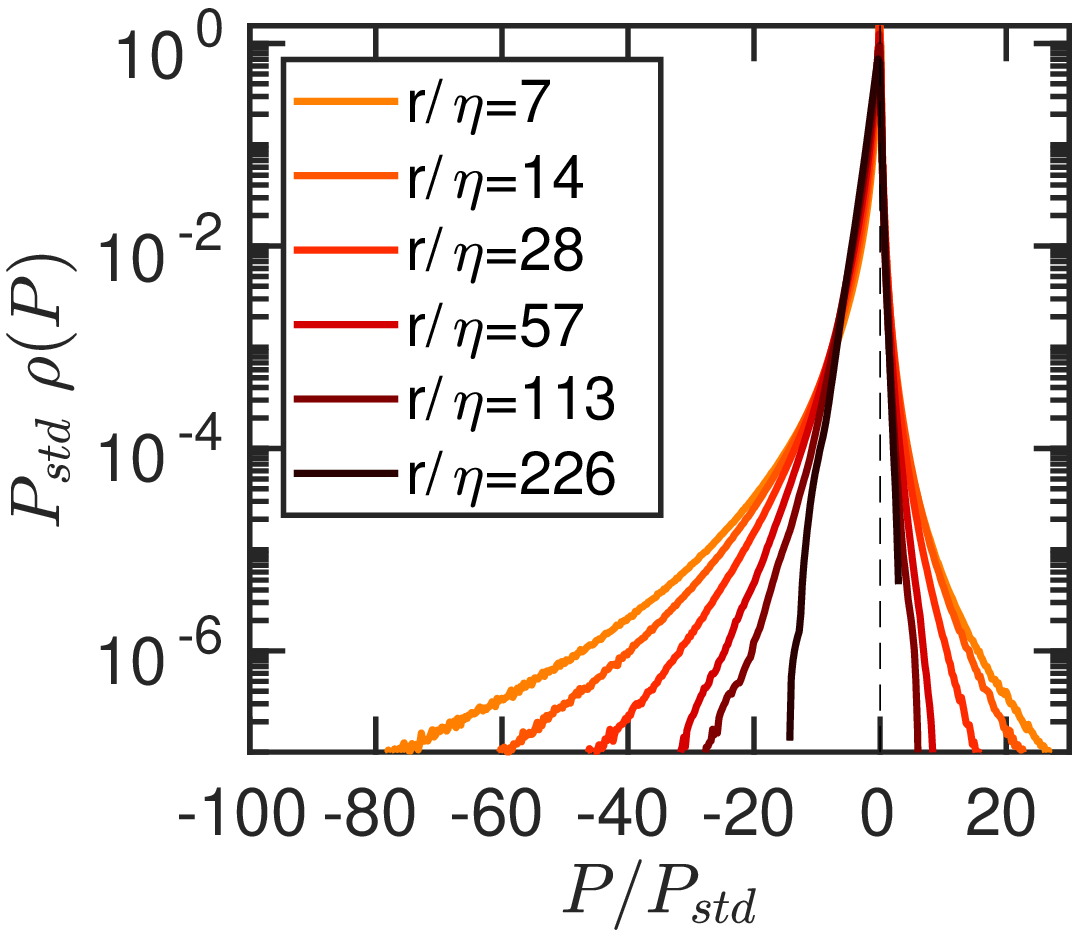}\hspace{1cm}
    \includegraphics[height=4.2cm]{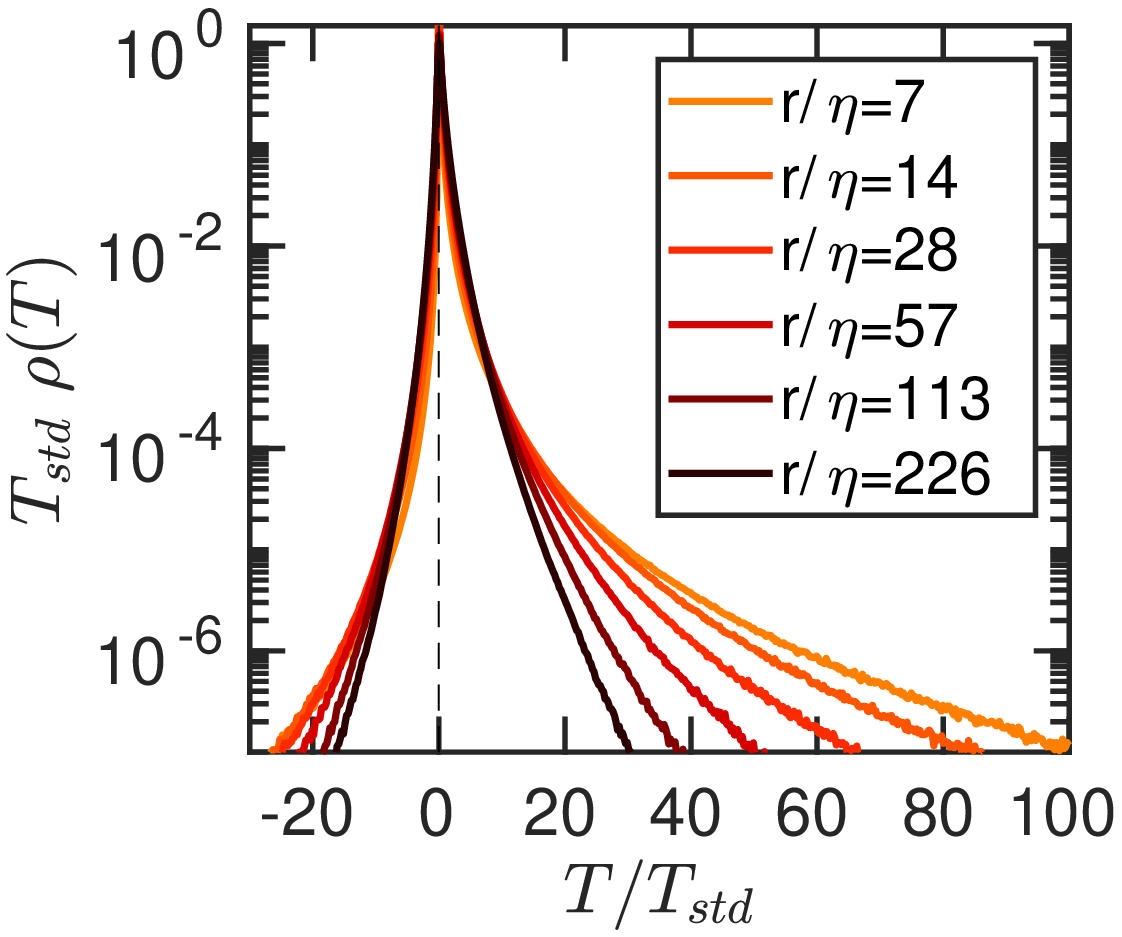}\\
    \hspace{-5.1cm} (a) \hspace{5.5cm} (b)
    \caption{Probability density function $\rho$ of $P$ (a) and $T$
      (b), normalized by the respective standard deviations $P_{std}$
      and $T_{std}$. Case HIT.\label{fig:pdfs_T_P_gauss_all_r}}
  \end{center}
\end{figure}

The probability density function (PDF) of $P$ is known to be
negatively skewed and to depend on scale and filter type
\citep{Piomelli1981, Aoyama}.  The PDF of $T$ has not been documented,
and it is reported here along with that of $P$
(\Figref{fig:pdfs_T_P_gauss_all_r}). The skewness of $T$ is of
opposite sign to that of $P$, which is perhaps unsurprising since one
expects extreme negative $P$ events to have an extreme positive $T$
counterpart. What is remarkable in our view concerns the tails of
the PDFs in the backscatter region --- i.e., $P>0$ and
$T<0$. While the tails of $P$ for large $r$ suffer a drastic change,
also observed by \cite{Aoyama}, the tails in the $T<0$ side are less
affected by the changes in $r$, providing a rather different view of
backscatter than that obtained through $P$. The stronger
$r$ dependence of the shape of the PDFs of $P$ compared to those of
$T$ explains the major departure of the aspect of the flow structures
observed in \Figref{fig:structures}(d) from those in
\Figref{fig:structures}(a-c).

\section{Comparison between $T$ and $P$ for different filter shapes}
\label{sec:comparisons_filter}

We now focus on a single flow and test different filter
definitions. Taking case HIT, we filter the velocity field following
three filters defined in Fourier space as
\begin{eqnarray}
\text{Gaussian:}\hspace{1cm} \tilde{G}(\boldsymbol{k})&=&\exp\left[-\left(r\boldsymbol{k} \right)^{2}/40 \right], \label{eqn:def_gauss2}\\
\text{Top-hat:}\hspace{1cm} \tilde{G}(\boldsymbol{k})&=&3\left[\sin\left(\zeta\right) - \zeta\cos\left(\zeta \right) \right]/\zeta^{3}, \label{eqn:def_tophat}\\
\text{Cutoff:}\hspace{1cm} \tilde{G}(\boldsymbol{k})&=& H\left( \frac{\pi}{r}-|\boldsymbol{k}| \right), \label{eqn:def_ksharp}
\end{eqnarray}
where $\zeta = |\boldsymbol{k}| r/2$ and $H$ is the Heaviside step
function. These definitions are adjusted so that the Gaussian and
top-hat filters have matched second moments
\citep{borue_orszag_1998}. It is practical to repeat \Eqnref{eqn:filt_k_def} here with 
\Eqnref{eqn:def_gauss2} for the sake of clarity. While the cutoff
filter is sharp in spectral space and oscillates in real space, the
top-hat has the opposite behavior with a sharp behavior in real space
and oscillations in Fourier space. The Gaussian filter is free from oscillations in
both spaces.
\begin{figure}
  \vspace{0.2cm}
  \begin{center}
    \includegraphics[width=0.4\textwidth]{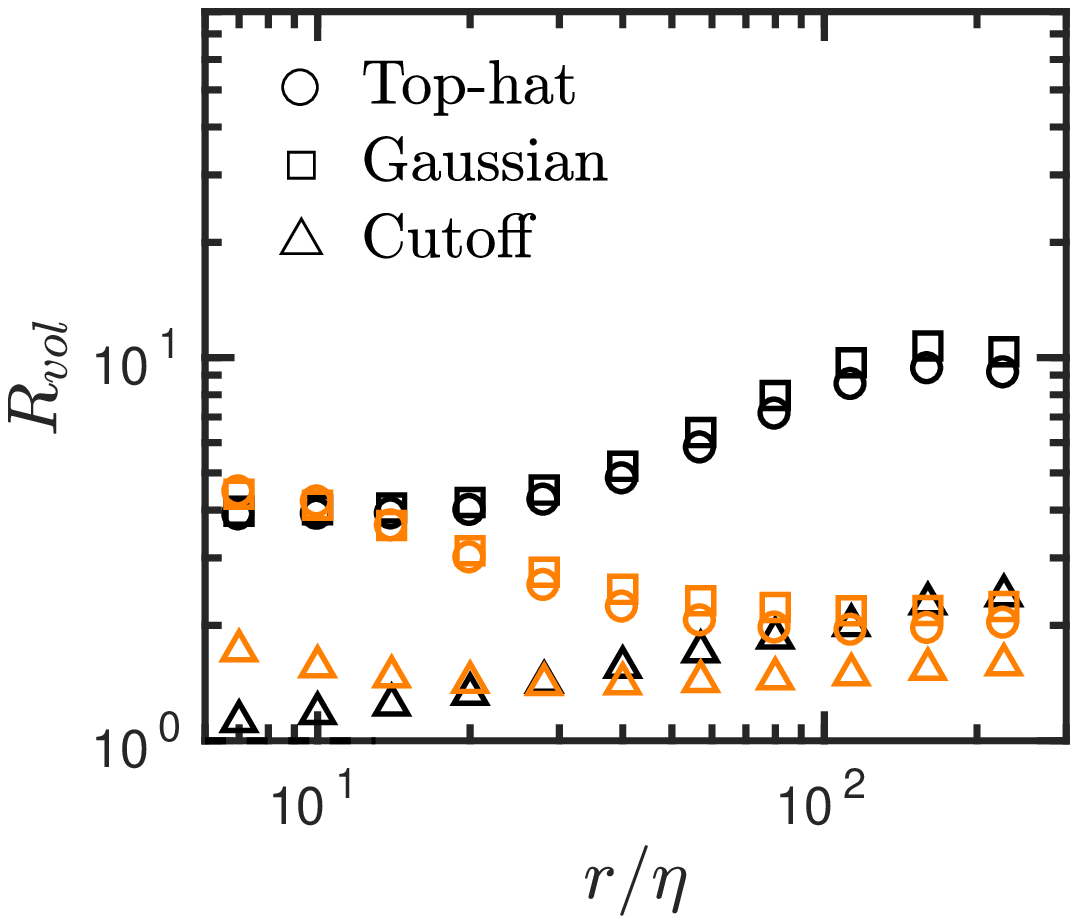}\hspace{.5cm}
    \includegraphics[width=0.4\textwidth]{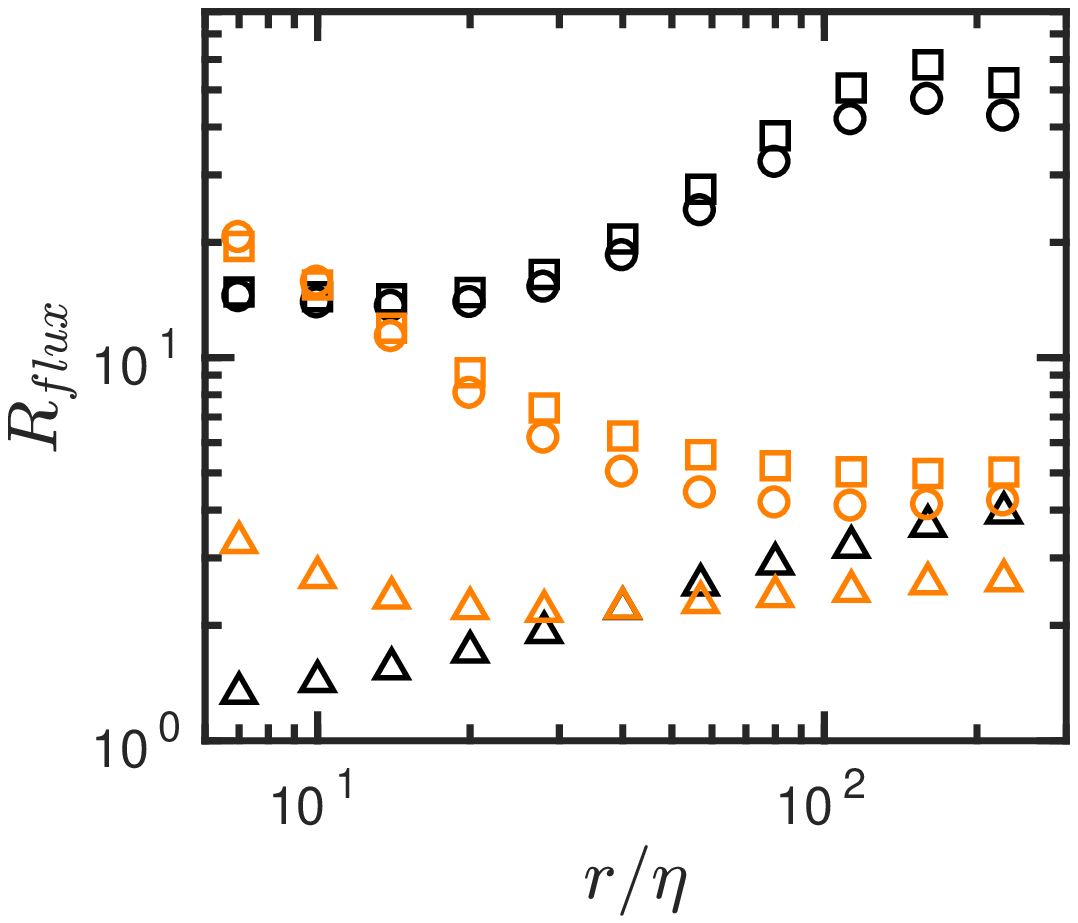}\\
    \hspace{-5.1cm} (a) \hspace{5.5cm} (b)
    \caption{Ratios of direct (forward) to inverse (backward)
      cascade based on $T$ (orange) and $P$ (black). For $T$, forward and
      backward cascade correspond to $T>0$ and $T<0$,
      respectively. $P<0$ and $P>0$ are taken as forward and backward
      cascade, respectively. Case HIT. (a) (flow volume forward) / (flow volume
      backward). (b) (energy flux forward) / (energy flux
      backward).\label{fig:cascade_ratios_filts}}
  \end{center}
\end{figure}
%

The ratios of forward to backward cascade based on volume fractions
and on energy fluxes are shown in
\Figref{fig:cascade_ratios_filts}. For a given ISET term, the ratios
based on the top-hat and Gaussian filters are in good agreement with
each other. The cutoff filter gives a different set of ratios
regardless of ISET term. The two filters that do not oscillate in real
space show cascade ratios which compare well across ISET terms for the
low-end range of $r$ values, but disagreement between $T$-
and $P$-based ratios grows for larger filter widths.  For $r$
larger than the viscous-dominated scales, the ratios based on $T$ vary
less with filter type than those based on $P$, with the cutoff filter
showing similar ratios to the other two filters for $T$-based ratios
and not so for $P$-based ratios.  

\begin{figure}
  \vspace{0.1cm}
  \begin{center}
    \includegraphics[height=4.2cm]{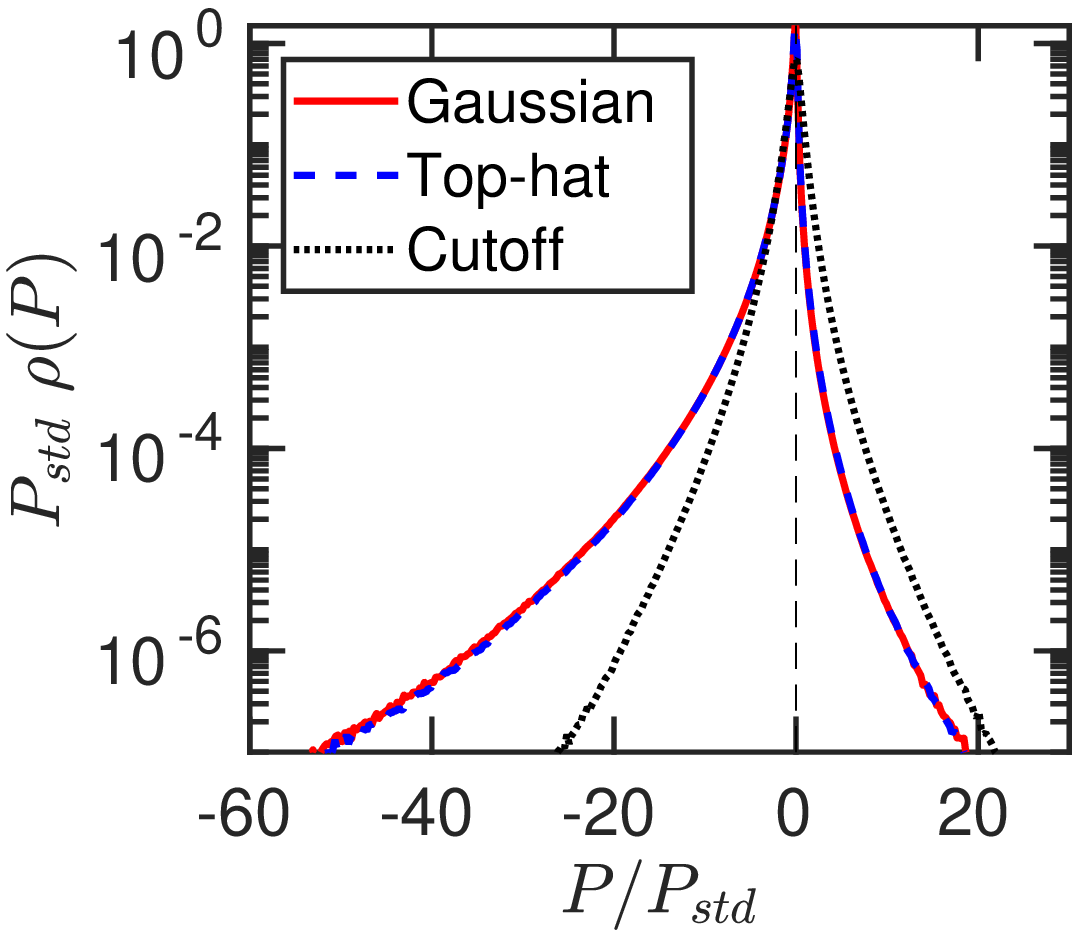}\hspace{1cm}
    \includegraphics[height=4.2cm]{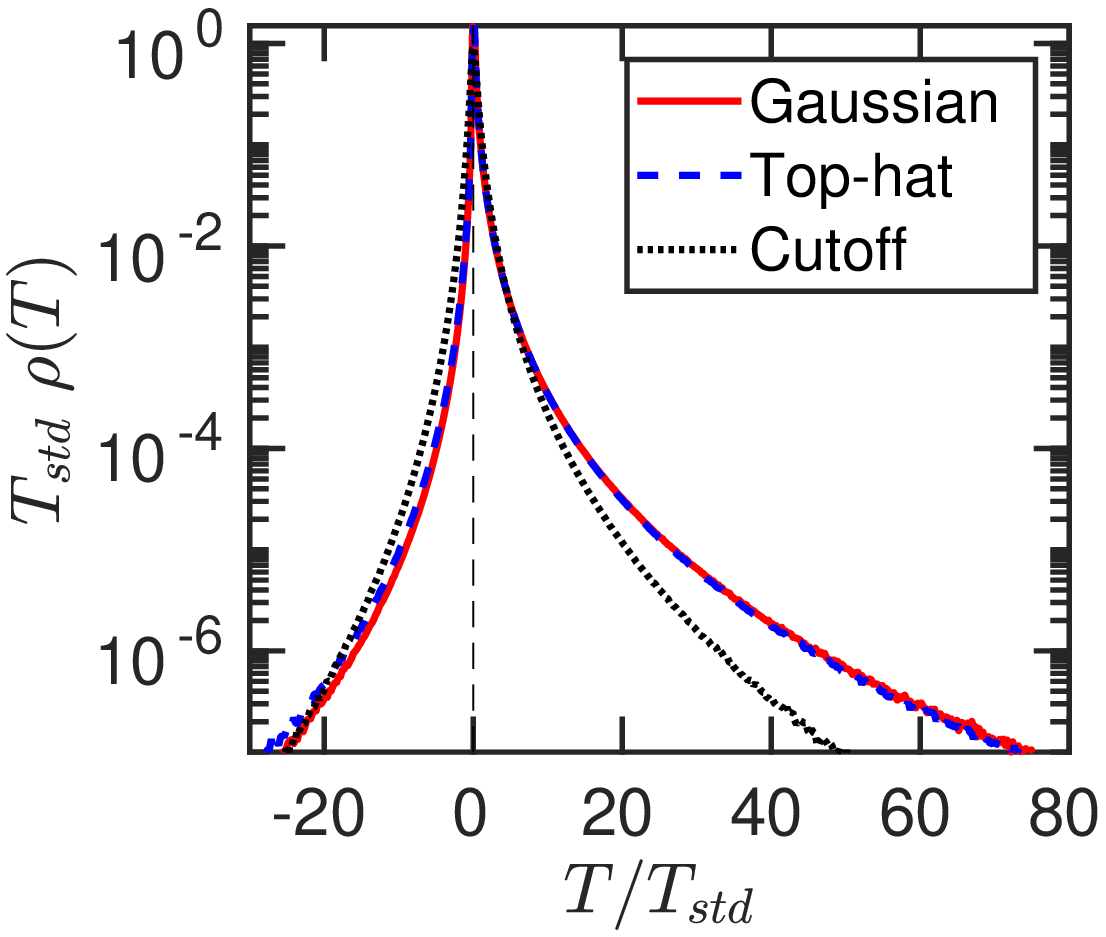}\\
    \hspace{-5.1cm} (a) \hspace{5.5cm} (b)\\
    \vspace{0.2cm}
    \includegraphics[height=4.2cm]{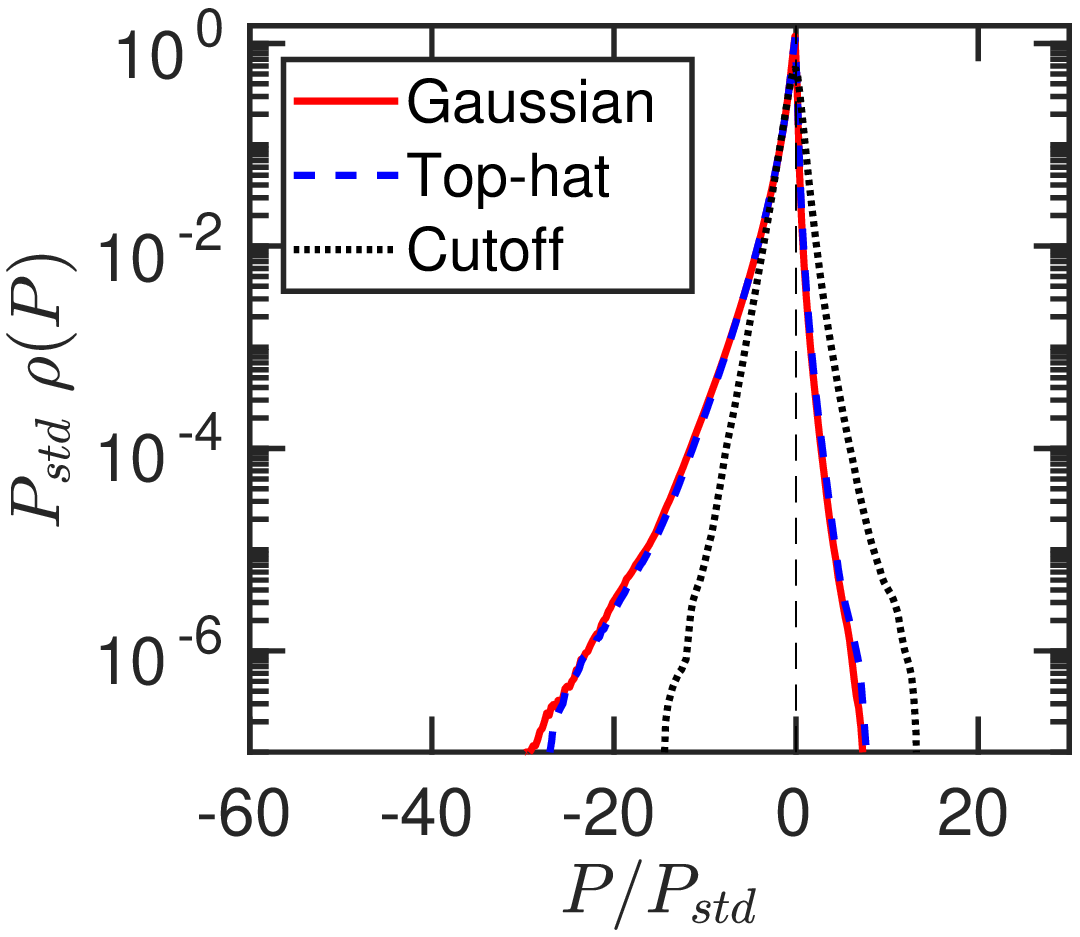}\hspace{1cm}
    \includegraphics[height=4.2cm]{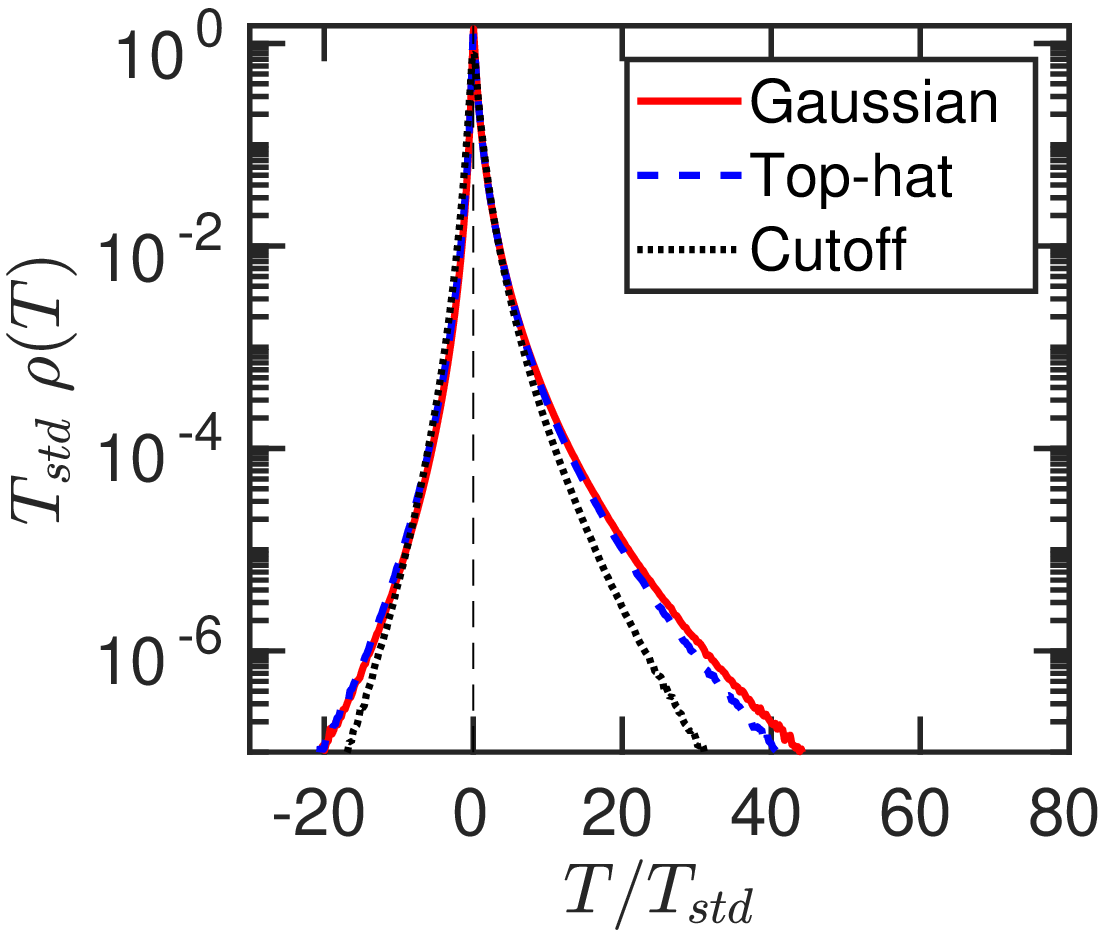}\\
    \hspace{-5.1cm} (c) \hspace{5.5cm} (d)\\
    \vspace{0.2cm}
    \includegraphics[height=4.2cm]{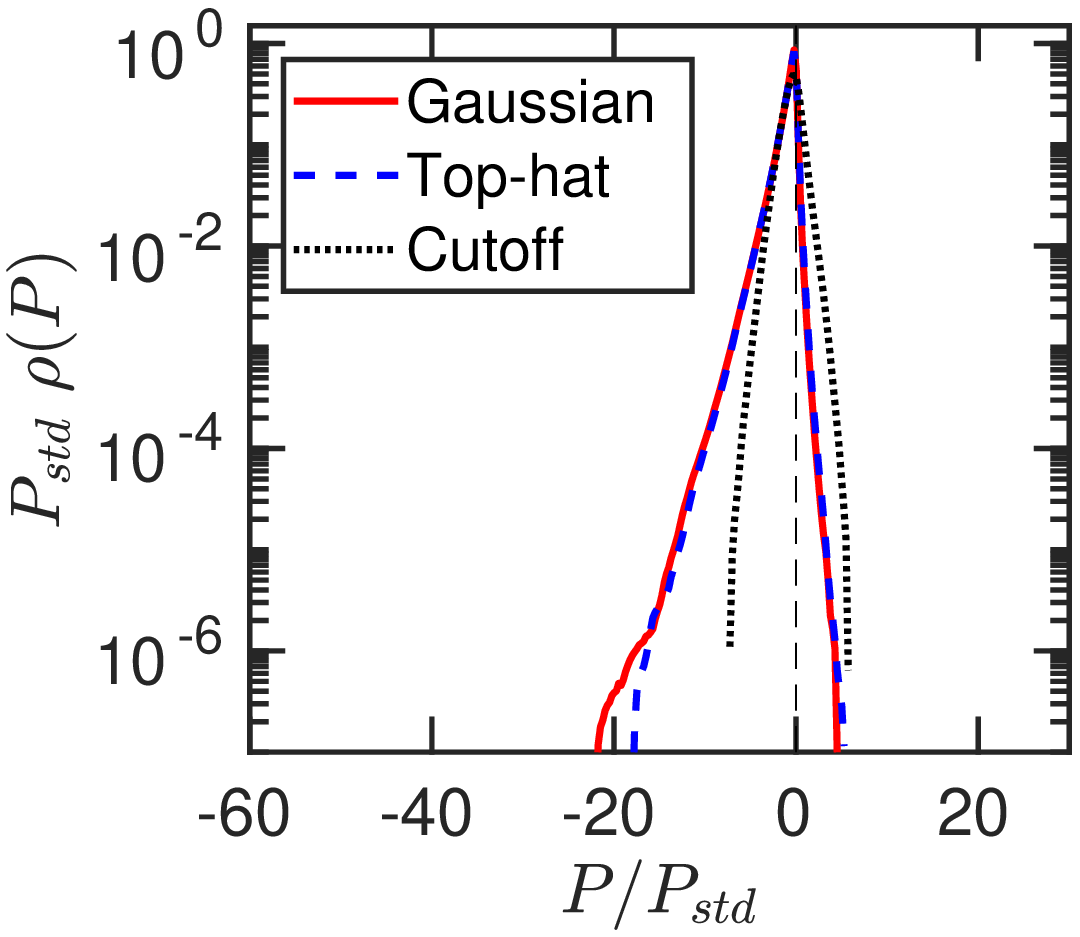}\hspace{1cm}
    \includegraphics[height=4.2cm]{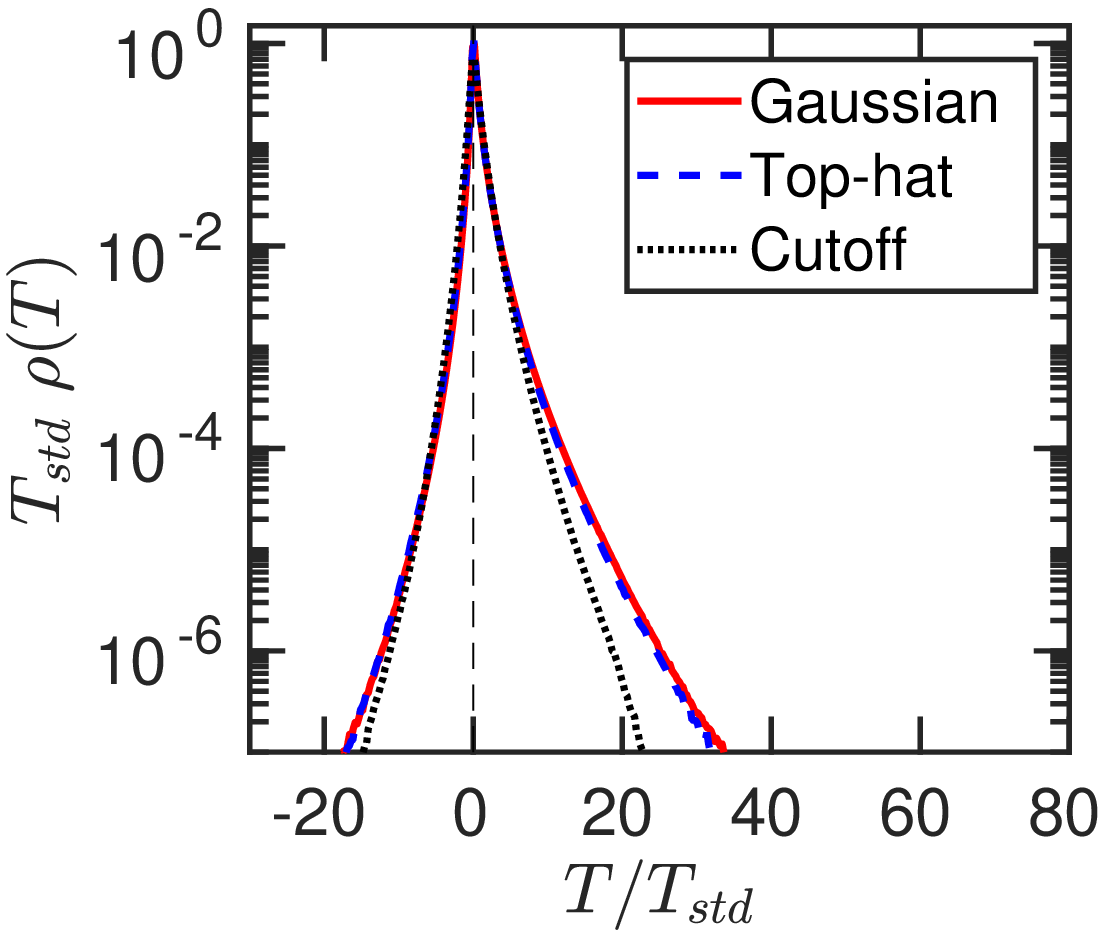}\\
    \hspace{-5.1cm} (e) \hspace{5.5cm} (f)
    \caption{Probability density function $\rho$ of $P$ (left) and $T$
      (right) for the three different filters defined in
      \Eqnrefs{eqn:def_gauss2}-(\ref{eqn:def_ksharp}), normalized by
      the respective standard deviations $T_{std}$ and $P_{std}$. Case HIT. (a)
      and (b): $r=20\eta$. (c) and (d): $r=80\eta$. (e) and (f):
      $r=160\eta$. \label{fig:pdfs_T_P_allfilts_3_r}}
  \end{center}
\end{figure}

The impact of the filter type on the
PDFs at three different scales is illustrated in
\Figref{fig:pdfs_T_P_allfilts_3_r}.  The PDFs on the right of
\Figref{fig:pdfs_T_P_allfilts_3_r} show that $T$ has distributions
based on the cutoff filter which significantly differ from those
obtained with the other two filters. However, the impact of the
cutoff filter on the PDFs of $P$ is even more striking at all three
values of $r$ considered.
%

\section{Conclusions}
\label{sec:conclusions}

The average transfer of kinetic energy from large to small flow motions
has been the cornerstone of most theories and models of the turbulence cascade
since the 1940s. Yet predicting the spatial structure of the kinetic energy 
transfer by going beyond the average
tendency and tackling the point-to-point description remains an
outstanding challenge in fluid mechanics. Despite the broad knowledge
acquired in recent years, advances in the characterization of the
energy cascade have been hampered by the ambiguity and disparity in the
tools and quantities to be considered.

Very often, the energy cascade has been studied via the filtered
equations of motion leading to the term
$P=\tau_{ij}\overline{S}_{ij}$
as the main figure of merit. Here, we have highlighted that the particular choice
of $P$ is, to some extent, arbitrary and that alternative definitions of
ISET are possible while still preserving the same
physical meaning ascribed to $P$. We have argued that
distinct ISET terms appear depending on the arrangement of
the right-hand side of the filtered kinetic energy equation. We have
illustrated this using as an example the term $P_2 = \overline{u}_i
\partial_j \tau_{ij}$.

However, while alternative formulations of $P$ exist, they do not
comply with our requirements for a physically sound ISET term.
The deficiencies can be traced back to i) the Galilean invariance of the rate of
energy change as a whole, ii) the Galilean invariance of ISET and other terms 
in an equation, and iii) the
remaining filter-subfilter coupling inside the spatial transport term
leading to erroneous interpretations as energy flux in space.
We have shown that $P$, despite its
Galilean invariance, is within an equation where the total rate of energy change
depends on the inertial frame of reference and filter-subfilter coupling is
contained in the spatial transport part of \Eqnref{eqn:kin_e_filt}. 
$P_2$ lacks Galilean invariance.

To address the shortcomings exposed above, we have proposed to study
the energy cascade from the point of view of the SFS
kinetic energy equation. We have shown that, by a judicious
arrangement of the terms in the equations, it
is possible to formulate an ISET term which i)
belongs to an equation that conserves the value of the rate of change of
energy under Galilean transformations, ii) is composed of terms
that are all individually Galilean invariant, 
and iii) leaves no remaining filter-subfilter coupling inside the spatial transport term.
This new term is given by $T = u'_{i} \partial_{j}
\left(\tau_{ij}\right)-u'_{i}u'_{j} \overline{S}_{ij}$.

We have studied the properties of $T$ and $P$ as markers to describe
the energy cascade. To that end, we analyzed data from DNS of
three flow configurations: HST, HIT, and turbulent channel flow. 
The ratios of forward to backward cascade based on $T$ are roughly
independent of the flow despite the very different large scales. This contrasts
with ratios based on $P$, which exhibit strong flow and filter width dependence.
The statistical properties of $T$, such as the amount of forward cascade and
backscatter, show milder dependence on the filter type than those
of $P$, at least within the range of filter types considered here. 
In conclusion, our results suggest that $T$ portrays
a picture of the energy cascade that is much closer to self-similar and universal
across flows and for a wider range of scales, in contrast to the picture
provided by the traditional $P$.

The present study is centered about improvements in the physical
understanding of the turbulent cascade rather than in practical
model implementation. We leave for future studies the connection
between $T$ and improved SFS models for LES.

\vspace{-0.7cm}
\setcounter{equation}{0}
\renewcommand\theequation{A.\arabic{equation}}
\section*{Appendix A. Galilean invariance of filtered velocities}
\label{sec:appendix}

The filter function $G$ in \Eqnref{eqn:decomp_def} satisfies the
normalization condition
\begin{equation}
\int G(r_{i}) dr_{i}=1 \label{eqn:norm_filter}.
\end{equation}
The Galilean transformation applied to the filtered velocity yields
\begin{equation}
\nonumber \overline{u}_{i} = \int u_{i}(x_{j}-r_{j}) G(r_{j}) dr_{j} = \int \hat{u}_{i}(x_{j}-r_{j})G(r_{j})dr_{j} + \int V_{i}G(r_{j})dr_{j},
\end{equation}
\begin{equation}
 = \int \hat{u}_{i}(x_{j}-r_{j})G(r_{j})dr_{j} + V_{i}\int G(r_{j}) dr_{j} = \overline{\hat{u}}_{i} + V_{i} \label{eqn:xxx},
\end{equation}
so that the filtered velocity depends on the frame velocity
$V_{i}$. For this reason, the terms $P_{2}$ in \Eqnref{eqn:kin_e_filt2}
and $\overline{u}_{i}\partial_{j}(u'_{i}u'_{j})$, which appear in
alternative ways of writing the right-hand side of
\Eqnrefs{eqn:kin_e_resA} and (\ref{eqn:kin_e_resD}), are not Galilean
invariant. On the contrary, the residual velocity $u'_{i}$ is given
by
\begin{equation}
  u'_{i}=u_{i}-\overline{u}_{i}=\hat{u}_{i}+V_{i}-\overline{u}_{i} =
  \hat{u}_{i}+V_{i}-\overline{\hat{u}}_{i} - V_{i} = \hat{u}_{i}-\overline{\hat{u}}_{i}=\hat{u}'_{i},
\end{equation}
which does not depend on the frame of reference. The residual velocity
in the transformed frame appears as the difference between the total
and the filtered velocities in the transformed frame, leaving the
relation between the three unchanged by the transformation. Applying
the transformation to the left-hand side of
\Eqnrefs{eqn:kin_e_resA}-(\ref{eqn:kin_e_resD}), we obtain
\begin{equation}
\left( \partial_{t}+ \overline{u}_{j}\partial_{j}\right) \frac{1}{2}u'_{i}u'_{i}=\left[ 
\hat{\partial}_{t}-V_{j}\hat{\partial}_{j}+\left(\overline{\hat{u}}_{j}+V_{j}\right) \hat{\partial}_{j}\right]\frac{1}{2}u'_{i}u'_{i}=
\left(\hat{\partial}_{t}+\overline{\hat{u}}_{j} \hat{\partial}_{j}\right)\frac{1}{2}\hat{u}'_{i}\hat{u}'_{i},
\end{equation}
so that the left-hand side conserves its value across Galilean frames
of reference. The SFS stress tensor is also Galilean invariant, as
shown below
\begin{eqnarray}
\nonumber \tau_{ij} &=& \overline{u_{i}u_{j}}-\overline{u}_{i}\overline{u}_{j}=\overline{(\hat{u}_{i}+V_{i})(\hat{u}_{j}+V_{j})}-\overline{\hat{u}_{i}+V_{i}}~\overline{\hat{u}_{j}+V_{j}},\\
\nonumber &=&\overline{\hat{u}_{i}\hat{u}_{j}}+\overline{\hat{u}_{i}V_{j}}+\overline{V_{i}\hat{u}_{j}}+\overline{V_{i}V_{j}}-\overline{\hat{u}}_{i}\overline{\hat{u}}_{j}-\overline{\hat{u}}_{i}\overline{V}_{j}-\overline{V}_{i}\overline{\hat{u}}_{j}
-\overline{V}_{i}\overline{V}_{j},\\
\nonumber &=&\overline{\hat{u}_{i}\hat{u}_{j}}+\overline{\hat{u}}_{i}V_{j}+V_{i}\overline{\hat{u}}_{j}+V_{i}V_{j}-\overline{\hat{u}}_{i}\overline{\hat{u}}_{j}-\overline{\hat{u}}_{i}V_{j}-V_{i}\overline{\hat{u}}_{j}
-V_{i}V_{j}=\overline{\hat{u}_{i}\hat{u}_{j}}-\overline{\hat{u}}_{i}\overline{\hat{u}}_{j} = \hat{\tau}_{ij}.
\end{eqnarray}
This explains why, as stated in Section \ref{sec:equations_res}, compliance
with a) also holds for the evolution equation of $\tau_{ii}$ written below
\begin{eqnarray}
\nonumber &&\left( \partial_{t}+\overline{u}_{j}\partial_{j} \right)\frac{1}{2} \tau_{ii} = 
                                                                                    -\tau_{ij}\overline{S}_{ij}
                                                                                    + 2\nu \overline{S}_{ij}\overline{S}_{ij} - 2\nu \overline{S_{ij}S_{ij}}+\overline{u_{i}F_{i}}-
  \overline{u}_{i}\overline{F}_{i}\\
&&-\partial_{j} \left( \overline{u_{j}p}-\overline{u}_{j}\overline{p}
              +2\nu\overline{u}_{i}\overline{S}_{ij}+2\nu\overline{u_{i}S_{ij}}+ 
              \frac{1}{2}\overline{u_{i}u_{i}u_{j}}-\frac{1}{2}\overline{u_{i}u_{i}}\overline{u}_{j}-\overline{u}_{i}\tau_{ij}\right). 
  \label{eqn:tau_ii}
\end{eqnarray}
To prove that \Eqnref{eqn:tau_ii} does not comply with b), we apply a Galilean transformation to
the last three terms in the divergence:
\begin{eqnarray}
\overline{u_{i}u_{i}u}_{j} &=& \overline{\hat{u}_{i}\hat{u}_{i}\hat{u}}_{j} + 2V_{i}\overline{\hat{u}_{i}\hat{u}}_{j} + V^{2}_{i}\overline{\hat{u}}_{j} + V_{j}\overline{\hat{u}_{i}\hat{u}}_{i}
+2V_{i}V_{j}\overline{\hat{u}}_{i} + V^{2}_{i}V_{j},\\
\overline{u_{i}u}_{i}\overline{u}_{j} &=& \overline{\hat{u}_{i}\hat{u}}_{i}\overline{\hat{u}}_{j} + 2V_{i}\overline{ \hat{u}}_{i}\overline{\hat{u}}_{j} + V^{2}_{i}\overline{\hat{u}}_{j}
+ V_{j}\overline{\hat{u}_{i}\hat{u}}_{i} +2V_{i}V_{j}\overline{\hat{u}}_{i} + V^{2}_{i}V_{j},\\
\overline{u}_{i}\tau_{ij} &=& \overline{\hat{u}}_{i} \overline{\hat{u}_{i}\hat{u}}_{j} - \overline{\hat{u}}_{i}\overline{\hat{u}}_{i}\overline{\hat{u}}_{j} + 
                              V_{i}\overline{\hat{u}_{i}\hat{u}}_{j} - V_{i}\overline{\hat{u}}_{i}\overline{\hat{u}}_{j} = \overline{\hat{u}}_{i} \hat{\tau}_{ij}
+V_{i}\overline{\hat{u}_{i}\hat{u}}_{j} - V_{i}\overline{\hat{u}}_{i}\overline{\hat{u}}_{j}.
\end{eqnarray}
None of these three terms is individually Galilean invariant, because in general $\overline{u_{i}u_{i}u}_{j}\neq\overline{\hat{u}_{i}\hat{u}_{i}\hat{u}}_{j}$,
$\overline{u_{i}u}_{i}\overline{u}_{j}\neq \overline{\hat{u}_{i}\hat{u}}_{i}\overline{\hat{u}}_{j}$, and
$\overline{u}_{i}\tau_{ij} \neq \overline{\hat{u}}_{i} \hat{\tau}_{ij}$.
Only the linear combination found in
the divergence of \Eqnref{eqn:tau_ii} leads to a Galilean-invariant expression
\begin{equation}
  \frac{1}{2}\overline{u_{i}u_{i}u}_{j} - \frac{1}{2}\overline{u_{i}u}_{i}\overline{u}_{j} - \overline{u}_{i}\tau_{ij}
= 
 \frac{1}{2}\overline{\hat{u}_{i}\hat{u}_{i}\hat{u}}_{j} - \frac{1}{2}\overline{\hat{u}_{i}\hat{u}}_{i}\overline{\hat{u}}_{j} - \overline{\hat{u}}_{i}\hat{\tau}_{ij},
\end{equation}
Since in general individual products do not have the same value on different inertial frames of reference, we face the problem that, depending on the inertial frame,
the relative importance of some products with respect to other products is not conserved. The situation is unsatisfactory
from a fundamental point of view, hence the reason for bringing requirement b) forward.
Finally, the presence of $\overline{u}_{i}\tau_{ij}$ inside the divergence in \Eqnref{eqn:tau_ii} implies a failure to comply with c).   


\vspace{-0.4cm}
\section*{Acknowledgments}

A.L.-D. acknowledges support from NASA under grant no. NNX15AU93A
and from ONR under grant no. N00014-16-S-BA10. J.I.C. acknowledges
funding from the Multiflow project of the European Research Council,
which financed this research while at the Technical University of
Madrid. We are grateful to Perry Johnson, Jane Bae, and Javier
Jim{\'e}nez for fruitful discussions.

\bibliographystyle{ctr}

\end{document}